\documentclass[12pt]{iopart}

\usepackage{iopams}

\usepackage{graphicx}
\usepackage{dcolumn}

\usepackage{bm}
\usepackage{color}
\usepackage{epstopdf}
\usepackage{soul}
\newtheorem{Proposition}{Proposition}

\begin{document}

\newcommand{\beq}{\begin{equation}}
\newcommand{\eeq}{\end{equation}}
\newcommand{\barr}{\begin{eqnarray}}
\newcommand{\earr}{\end{eqnarray}}

\newcommand{\RED}[1]{\textrm{\color{red}#1}}
\newcommand{\rev}[1]{{\color{red}#1}}
\newcommand{\REV}[1]{{\color{red}[[Saverio: #1]]}}
\newcommand{\BLUE}[1]{{\color{blue}[[Saverio: #1]]}}
\newcommand{\GREEN}[1]{{\color{green}[[DAREK: #1]]}}

\newcommand{\ket}[1]{|#1\rangle}
\newcommand{\bra}[1]{\langle#1|}

\newcommand{\bmsub}[1]{\mbox{\boldmath\scriptsize $#1$}}

\def\R{\mathbf{R}}
\def\C{\mathbf{C}}
\def\bra#1{\langle #1 |}
\def\ket#1{| #1 \rangle}
\def\sinc{\mathop{\text{sinc}}\nolimits}
\def\cV{\mathcal{V}}
\def\cH{\mathcal{H}}
\def\cT{\mathcal{T}}
\def\cM{\mathcal{M}}
\def\cN{\mathcal{N}}
\def\CW{\mathcal{W}}
\def\cA{\mathcal{A}}
\def\e{\mathrm{e}}
\def\ii{\mathrm{i}}
\def\d{\mathrm{d}}
\renewcommand{\Re}{\mathop{\text{Re}}\nolimits}
%\newcommand{\tr}{\mathop{\text{Tr}}\nolimits}

%DAREK
\newcommand{\MN}{M_N(\mathbb{C})}
\newcommand{\Mk}{M_k(\mathbb{C})}
\newcommand{\id}{\mbox{id}}
\newcommand{\ot}{{\,\otimes\,}}
\newcommand{{\Cd}}{{\mathbb{C}^d}}
\newcommand{\sbsigma}{{\mbox{\scriptsize \boldmath $\sigma$}}}
\newcommand{\sbalpha}{{\mbox{\scriptsize \boldmath $\alpha$}}}
\newcommand{\sbbeta}{{\mbox{\scriptsize \boldmath $\beta$}}}
\newcommand{\sba}{{\mbox{\scriptsize \boldmath $a$}}}
\newcommand{\sbb}{{\mbox{\scriptsize \boldmath $b$}}}
\newcommand{\sbmu}{{\mbox{\scriptsize \boldmath $\mu$}}}
\newcommand{\sbnu}{{\mbox{\scriptsize \boldmath $\nu$}}}
\def\oper{\mathbf{1}}
\def\<{\langle}
\def\>{\rangle}
\newtheorem{theorem}{Theorem}
\newtheorem{definition}{Definition}
\newtheorem{remark}{Remark}
%END  DAREK

\title{Dynamically contracted algebra of observables for dissipative quantum systems }

\author{Sahar Alipour$^1$, Dariusz Chru\'sci\'nski$^2$, Paolo Facchi$^{3,4}$, Giuseppe Marmo$^{5,6}$, Saverio Pascazio$^{3,4}$, and Ali T. Rezakhani$^1$}
\address{$^{1}$Department of Physics, Sharif University of Technology, Tehran 14588, Iran}
\address{$^2$Institute of Physics, Nicolaus Copernicus University, Grudzi\c{a}dzka 5/7, 87--100 Toru\'n, Poland}
\address{$^{3}$Dipartimento di Fisica and MECENAS, Universit\`a di Bari, I-70126  Bari, Italy}
\address{$^{4}$INFN, Sezione di Bari, I-70126 Bari, Italy}
\address{$^{5}$Dipartimento di Scienze Fisiche and MECENAS, Universit\`a di Napoli ``Federico II", I-80126 Napoli, Italy}
\address{$^{5}$INFN, Sezione di Napoli, I-80126  Napoli, Italy}
\date{\today}

\begin{abstract}
A method is discussed to analyze the dynamics of a dissipative quantum system. The method hinges upon the definition of an alternative (time-dependent) product among the observables of the system. In the long time limit this yields a contracted algebra. This contraction irreversibly affects some of the quantum features of the dissipative system.
\end{abstract}


\maketitle

\section{Introduction}
\label{sec:intro}

Coherence is at the heart of the most genuinely non-classical aspects of a quantum-mechanical system.
A dissipative process provokes a deterioration of the quantum features and makes the system more ``classical". For instance, coherence between the two branch waves in a double-slit experiment is a fundamental requisite to observe interference. A decohering environment affects the ability of a quantum particle to interfere, making its behavior more ballistic and classical.

It is natural to expect that the features of the dissipative dynamics heavily influence the quantum-to-classical transition. Intuitively, one would expect that during a dissipative process it becomes increasingly difficult to unearth some of the non-classical aspects of a quantum system. These limitations should be reflected in the measurements that one can perform on the system and therefore should modify its algebra of observables.

In this article we shall tackle this problem for Markovian quantum systems, whose dynamics can be formulated according to Gorini, Kossakowski, Sudarshan~\cite{GKS} and Lindblad (GKSL)~\cite{lind}, within the framework of quantum dynamical semigroups~\cite{Alicki}. Quantum dissipative systems~\cite{Weiss,Breuer} have a wide range of applications and are of paramount importance in quantum technologies and quantum information~\cite{QIT}.

The objective of this article is to discuss and elaborate on a mathematical mechanism that provokes a deformation and eventually a contraction of the algebra of the observables of a dissipative system \cite{CFMP}. This mechanism hinges upon an alternative definition of the product among observables. We will study the effects of the contraction on the full associative algebra of operators and will see that some (initially nonvanishing) commutators eventually vanish as a consequence of dissipation: the associated observables become simultaneously measurable and in this sense the system becomes more ``classical". The approach we propose here bears analogies but also differences with the macroscopic and semi-classical limit~\cite{wigner,araki,jauch}. In both cases, the system is considered classical when some or all of its observables commute. On the other hand, we shall adopt here a pure algebraic standpoint, that does not involve an explicit macroscopic limit (although the presence of a bath, endowed with infinite degrees of freedom, is implicitly assumed when one writes a GKSL equation).

We shall see that the asymptotic state plays a key role in our description. Several features of the contracted algebra can be properly understood only if one computes the expectation value over the asymptotic state of the evolution. In addition, one needs to contract the full associative algebra of operators: following the Lie algebra alone is not sufficient to draw proper conclusions on the dynamics. This differs from the approach proposed in \cite{CFMP} and elaborates new facets of the contraction procedure.

This article is organized as follows. The alternative product and the general framework are introduced in Section~\ref{sec:intro2}. Section~\ref{sec:examples} is entirely devoted to some case studies, which help familiarize with the scheme and its physical consequences. A generic $n$-level system is analyzed in Section~\ref{sec:gener}. We conclude with some comments and an outlook in Section~\ref{sec:concl}.

\section{A new product}
\label{sec:intro2}

Let the algebra of observables of a quantum system belong to the Banach space $\mathcal{B}(\mathcal{H})$ of bounded linear operators defined on its Hilbert space $\mathcal{H}$. Assume that the dynamics be Markovian, so that the evolution equation for the density matrix $\varrho$ reads
\begin{equation}
\label{Msol}
\dot \varrho(t) = L [\varrho(t)]  ,
\end{equation}
whose solution is
\begin{equation}
\label{Msol2}
\varrho(t)  = e^{tL}[\varrho(0)] = \Lambda_t [\varrho(0)] \quad (t \geqslant 0).
\end{equation}
Here and henceforth a dot denotes $\mathrm{d}/\mathrm{d}t$.
The adjoint dynamical equation for an observable $A$ is defined through
\begin{equation}
\label{M2}
    \dot{A}(t) = L^\sharp [A(t)]  \quad \Longleftrightarrow \quad A(t) =  \Lambda_t^\sharp [A(0)]
    \quad (t \geqslant 0).
\end{equation}
The equivalence of the two descriptions hinges upon Dirac's prescription~\cite{Diracbook}
\begin{equation}
\label{SvsH}
\mathrm{Tr}[\varrho(t) A(0)] = \mathrm{Tr}[\varrho(0) A(t)],~~\forall \varrho,A
\end{equation}
that connects the Schr\"odinger and Heisenberg pictures.

We will focus on the effects of the adjoint evolution $\Lambda^\sharp_t$ on the product of observables. Let $\cA\subset \mathcal{B}(\mathcal{H})$ be the algebra of observables and $\{A_j\}$ a basis. The product ``$\circ$" between observables reads
\begin{equation}
\label{proddef}
A_i \circ A_j = \alpha_{ij}^{k} A_k,
\end{equation}
and is naturally defined as $(A\circ B)|\psi\rangle =A(B|\psi\rangle)$ $\forall |\psi\rangle\in\mathcal{H}$. This enables one to define the commutators (Lie product) through the structure constants $c$,
\begin{equation}
\label{Cstruct}
[A_i,A_j] = c_{ij}^{k} A_k, \qquad c_{ij}^{k} = \alpha_{ij}^{k} - \alpha_{ji}^{k},
\end{equation}
where $[A,B] = A\circ B - B \circ A$,
and the anticommutators (Jordan product), through
\begin{equation}
\label{Cantistruct}
\{A_i,A_j\} = s_{ij}^{k} A_k, \qquad s_{ij}^{k} = \alpha_{ij}^{k} + \alpha_{ji}^{k},
\end{equation}
where $\{A,B\}=A \circ B+B \circ A$.

In this article we focus on the time-dependent product~\cite{CGM,CFMP}
\begin{equation}
\label{prodpauli0}
A \circ_t B \equiv (\Lambda_t^\sharp)^{-1} \left[\Lambda_t^\sharp [A] \circ \Lambda_t^\sharp [B]\right], \quad \forall  A,B \in \cA ,
\end{equation}
that can be expressed as
\begin{equation}
\label{prodpauli}
A_i \circ_t A_j = \alpha_{ij}^{k}(t) A_k .
\label{tprod}
\end{equation}
Clearly, $\circ_{t=0}=\circ$ and $\alpha_{ij}^{k}(t=0)= \alpha_{ij}^{k}$.
We observe that the time-dependent product~(\ref{tprod}) yields an algebra homomorphism between $(\cA, \circ_t)$ and $(\Lambda_t^\sharp[\cA], \circ)$ $\forall t$, since $\Lambda_t^\sharp [A \circ_t B]=\Lambda_t^\sharp [A] \circ \Lambda_t^\sharp [B]$. It is evident that definition~(\ref{prodpauli}) depends on the existence of the inverse $ (\Lambda_t^\sharp)^{-1}$. Commutators
\begin{equation}
\label{commpauli}
[A_i , A_j]_t \equiv (\Lambda_t^\sharp)^{-1} \left(\left[\Lambda_t^\sharp (A_i) , \Lambda_t^\sharp (A_j)\right]\right) \equiv   c_{ij}^{k}(t) A_k
\end{equation}
and anticommutators
\begin{equation}
\label{anticommpauli}
\{A_i , A_j\}_t \equiv (\Lambda_t^\sharp)^{-1} \left(\left\{\Lambda_t^\sharp (A_i) , \Lambda_t^\sharp (A_j)\right\} \right)\equiv   s_{ij}^{k}(t) A_k
\end{equation}
are defined accordingly.

The $t \to \infty$ limit (\emph{if} exists) is naturally defined as
\begin{eqnarray}
\label{prodpauliinf}
A_i \circ_\infty A_j & \equiv & \lim_{t \to \infty} A_i \circ_t A_j = \alpha_{ij}^{k}(\infty) A_k,\\
\label{commpauliinf}
 [A_i , A_j]_\infty & \equiv & \lim_{t \to \infty} [A_i , A_j]_t = c_{ij}^{k}(\infty) A_k ,\\
 \label{anticommpauliinf}
  \{A_i , A_j\}_\infty & \equiv & \lim_{t \to \infty} \{A_i , A_j\}_t = s_{ij}^{k}(\infty) A_k .
\end{eqnarray}
In general, the above equations yield a \textit{contraction} $\cA_\infty$  of the original algebra $\cA$~\cite{IW,Sa,book:gilmore}. 

A few comments are in order. Note that some commutators will vanish in the $t \to \infty$ limit, so that the contracted algebra will always contain one or more Abelian subalgebras. In this sense, the system becomes more ``classical," in accord with physical intuition about dissipative dynamics. Observe also that the limiting product~(\ref{prodpauliinf}) captures (through the non-invertibility of the limiting map and the contraction of the algebra) a symptom of irreversibility that is not detected for any finite $t$. This can be viewed as a relic of the continuous interaction with an underlying infinite-dimensional dissipative bath. For finite-dimensional systems, one can observe different phenomena, such as collapses and revivals, that do not lead to any contraction.

In the following, the general idea that the contraction yields a method to discern the irreversibility of the dissipative dynamics will be tested on several examples. In particular, we shall illustrate the mechanism that affects the quantum features of the dissipative system.

\section{Case studies: qubits and quantum oscillators}
\label{sec:examples}

Let us illustrate the main ideas outlined in the previous section by looking at four case studies: (i) phase damping, (ii) energy damping, (iii) interaction with a thermal environment, and (iv) interaction with a squeezed environment. We shall see that in all cases the associative algebra of operators undergoes a contraction. These examples help elucidate a contraction through its various features.

\subsection{Phase damping}

\subsubsection{Phase damping of a qubit.}
\label{sec:first}

Our first paradigmatic example is the phase damping of a qubit~\cite{NP99}. We shall analyze this example with particular care, by solving it in different ways in order to pinpoint the origin of the contraction and its consequences.

The evolution of the density matrix of the qubit is described by Eq.~(\ref{Msol}) (we drop the explicit $t$-dependence here and in the following)
\begin{equation}
\label{qubit3}
\dot \varrho = L [\varrho] = - \frac \gamma 2 (\varrho- \sigma_3 \varrho \sigma_3),
\end{equation}
where $\gamma>0$, and $\sigma_\alpha \; (\alpha = 0,1,2,3)$ are the Pauli matrices (with $\sigma_0\equiv \mathbf{1}$). The adjoint dynamics for an observable $A$ is simply obtained by replacing any quantum jump operator with its adjoint (and eventually by changing $i$ in $-i$), so that
\begin{equation}
\label{M22}
 \dot{A} = L^\sharp [A]  = - \frac \gamma 2 (A- \sigma_3 A \sigma_3).
\end{equation}
This map is self-dual ($L = L^\sharp$ and $\Lambda=\Lambda^\sharp$) and yields
\begin{eqnarray}
& & \Lambda^\sharp_t [\sigma_{0,3}] = \Lambda^\sharp_\infty [\sigma_{0,3}] = \sigma_{0,3},
\label{commasympt0} \\
& & \Lambda^\sharp_t [\sigma_{1,2}] = e^{-\gamma t} \sigma_{1,2} \quad \to \quad \Lambda^\sharp_\infty [\sigma_{1,2}]=0.
\label{commasympt}
\end{eqnarray}
The associative product~(\ref{prodpauli}) yields
\begin{eqnarray}
\sigma_{0,3} \circ_t \sigma_{0,3} = \sigma_{0,3} \circ  \sigma_{0,3} \quad &\to& \quad \sigma_{0,3} \circ_{\infty} \sigma_{0,3} = \sigma_{0,3} \circ  \sigma_{0,3},
\\
\sigma_{1,2} \circ_t \sigma_{1,2} = e^{-2 \gamma t}  \sigma_{1,2} \circ \sigma_{1,2} \quad &\to& \quad
\sigma_{1,2} \circ_{\infty} \sigma_{1,2} = 0 \label{sigmasq},
\\
\sigma_{0,3} \circ_t \sigma_{1,2} = \sigma_{0,3} \circ \sigma_{1,2} \quad &\to& \quad \sigma_{0,3} \circ_{\infty} \sigma_{1,2} = \sigma_{0,3} \circ \sigma_{1,2}
\end{eqnarray}
and yields the following $\alpha(\infty)$ constants in Eq.~(\ref{prodpauliinf})
\begin{eqnarray}
\label{alfa2}
\alpha_{ij}^{k} (\infty) &=&
\left\{ \begin{array}{ll} \alpha_{ij}^{k} \ & \textrm{if}\; i\in\{0,3\}\; \textrm{or}\; j\in\{0,3\},
          \\ 0 \ & \textrm{if}\; i,j\in\{1,2\}. \end{array} \right. 
\end{eqnarray}
This leaves us with $\{\sigma_0 , \sigma_3 \}$, namely $\cA_\infty=\mathbb{C}^2$, the Abelian algebra of diagonal $2 \times 2$ matrices.

From the mathematical point of view, the explicit calculation of the contracted associative product is all one needs. In particular, from the $\alpha$ constants one can derive the structure constants characterizing both the Lie and Jordan algebras~(\ref{commpauliinf})-(\ref{anticommpauliinf}). However, as we shall see in this and in the following examples, the direct calculation of the Lie algebra may motivate interesting observations. In this example, direct computation would yield the Lie algebra
\begin{eqnarray}
\label{Masympt1}
& &\left[\sigma_1, \sigma_2\right]_t = 2 i  e^{-2\gamma t} \sigma_3
  \to
[\sigma_1,\sigma_2]_\infty=0, \\
& &\left[\sigma_2, \sigma_3\right]_t = 2 i  \sigma_1
 \to
[\sigma_2,\sigma_3]_\infty= 2 i  \sigma_1,
\\
& &\left[\sigma_3, \sigma_1\right]_t = 2 i  \sigma_2
 \to
[\sigma_3,\sigma_1]_\infty= 2 i  \sigma_2. \label{Masympt3}
\end{eqnarray}
whose asymptotic structure constants $c_{ij}^{k}(\infty)$ [Eq.~(\ref{anticommpauliinf})] are of course in agreement with the $\alpha(\infty)$ constants in Eq.~(\ref{alfa2}). According to Eqs.~(\ref{Masympt1})-(\ref{Masympt3}), the original $\mathfrak{su}(2)$ algebra contracts to the Euclidean algebra $\mathfrak{e}(2)$ of the isometries of the plane. However, one should also note that the latter is consistent with the Abelian algebra $\mathbb{C}^2$, if $\sigma_{1,2} \sim 0$, as dictated by Eq.~(\ref{sigmasq}). See also the following remarks.

The above picture is in accord with the asymptotic solution of Eq.~(\ref{qubit3}), that reads
\begin{equation}
\label{qubit3sol}
   \varrho = \frac{1}{2} (\sigma_0 + \bm x \cdot \bm\sigma) \stackrel{t \to \infty}{\longrightarrow}  \Lambda_\infty(\varrho)=\varrho_\infty = \frac{1}{2} (\sigma_0 + x_3 \sigma_3),
\end{equation}
$\bm x$ being a vector in the unit 3-dimensional ball, $|\bm x| \leqslant 1$.
All the preceding equations that involve operators in the $t \to \infty$ limit must be understood in the weak sense, according to Eq.~(\ref{SvsH}). For example, the expectation values of $\sigma_1$ and $\sigma_2$ on the asymptotic state~(\ref{qubit3sol}) vanish: as time goes by, it becomes increasingly difficult to measure the coherence (off-diagonal operators) between the two states of the qubit. In the $t\to\infty$ limit, coherence is fully lost. On the other hand, the expectation value of $\sigma_3$ does not vanish and the only nontrivial observables are the populations.
We note also that, on the asymptotic state~(\ref{qubit3sol})
\begin{equation}
(\Delta \sigma_{1,2})^2 = \langle (\sigma_{1,2})^2 \rangle - \langle \sigma_{1,2} \rangle^2 = 0-0 =0,
\end{equation}
where $(\sigma_{1,2})^2 = \sigma_{1,2} \circ_{\infty} \sigma_{1,2}$, whereas
\begin{equation}
(\Delta \sigma_{3})^2 = 1 - \langle \sigma_{3} \rangle^2 .
\end{equation}
The interpretation is  consistent: $\sigma_1$ and $\sigma_2$ weakly vanish in the asymptotic state~(\ref{qubit3sol}) and the only nontrivial observable besides $\mathbf{1}$ is $\sigma_3$. Thus $\cA_\infty=\mathbb{C}^2$.

One can obtain a deeper insight into the contraction of the algebra described above by looking at the problem from a wider perspective. The GKSL equation~(\ref{qubit3}) can be obtained from the following Hamiltonian:
\begin{equation}
H=\mathbf{1}\otimes\int_{\mathbb{R}}\omega a_\omega^\dag a_\omega \mathrm{d}\omega
+\frac{\Gamma}{2} \sigma_3 \otimes\int_{\mathbb{R}} (a_\omega+a_\omega^\dag)\mathrm{d}\omega,
\end{equation}
where $\Gamma = \sqrt{\gamma/2\pi}$, and $a_\omega$ ($a_\omega^\dag$) are the bosonic annihilation (creation) operators of the bath. The usual procedure~\cite{Davies,GZ} is to expand up to the second order in $\Gamma$ and trace out the bath degrees of freedom in order to obtain an evolution equation for the density matrix of the system. We shall, however, take in the following a different route.

The solutions to the Heisenberg operators read
\begin{eqnarray}
(\sigma_3\otimes \mathbf{1})(t) &= e^{iHt} (\sigma_3\otimes \mathbf{1}) e^{-iHt} =  \sigma_3\otimes\mathbf{1},
\label{HO1} \\
(\sigma_+\otimes\mathbf{1})(t) &= e^{iHt}(\sigma_+\otimes \mathbf{1}) e^{-iHt}  \nonumber \\
& =  \sigma_+ \otimes
\exp \left[ \Gamma \int \left(
 \frac{1-e^{-i\omega t}}{\omega} a_\omega-
\frac{1-e^{i\omega t}}{\omega} a_\omega^\dag
\right) \mathrm{d}\omega \right],
\label{HO2} \\
(\mathbf{1}\otimes a_\omega)(t) &=e^{iHt}(\mathbf{1}\otimes a_\omega) e^{-iHt}  =
e^{-i\omega t}  \mathbf{1}\otimes a_\omega - \frac{\Gamma}{2} \frac{1-e^{-i\omega t}}{\omega} \sigma_3\otimes \mathbf{1},
\label{HO3}
\end{eqnarray}
where $\sigma_\pm \equiv (\sigma_1 \pm i \sigma_2)/2$.

Clearly, the Schr\"odinger operators of the system in Eqs.~(\ref{HO1})-(\ref{HO2}) evolve into the Heisenberg operators that contain contributions of the bath. The idea is to identify the evolved operators of the system under the adjoint evolution equation~(\ref{M2}),~(\ref{M22}), with the trace  of the full Heisenberg operators over the ground state $|0\rangle$ of the bath:
\begin{equation}
\label{identifica}
A(t) = \Lambda_t^\sharp [A(0)] \equiv \langle (A \otimes \mathbf{1})(t) \rangle_{\rm{bath}} =
\tr_{\mathrm{bath}} \left\{(A\otimes\mathbf{1})(t) (\mathbf{1}\otimes |0\rangle\langle 0|)\right\}.
\end{equation}
We thus obtain
\begin{equation}
\label{identsig3}
\langle (\sigma_3 \otimes \mathbf{1})(t) \rangle_{\rm{bath}} = \sigma_3
\end{equation}
and 
\begin{eqnarray}
\label{identsim+}
\langle (\sigma_+\otimes \mathbf{1})(t) \rangle_{\rm{bath}} &=
   \langle 0|
\exp \left[\Gamma \int  \left(
 \frac{1-e^{-i\omega t}}{\omega}a_\omega-
\frac{1-e^{i\omega t}}{\omega}a_\omega^\dag
\right) \mathrm{d}\omega\right]
| 0 \rangle \, \sigma_+
\nonumber \\
&=
 \left( \Pi_\omega \langle 0|
\alpha_\omega \rangle\right) \sigma_+
\nonumber \\
&=
  \left(\Pi_\omega e^{-|\alpha_\omega|^2/2}\right) \sigma_+
\nonumber \\
&=
  \exp \left[ - 2 \Gamma^2 \int d \omega \frac{\sin^2(\omega t)}{\omega^2} \right] \sigma_+
\nonumber \\
&=
e^{- 2 \pi \Gamma^2 t} \sigma_+
\nonumber \\
&=
e^{- \gamma t} \sigma_+ ,
\end{eqnarray}
$|\alpha_\omega \rangle$ being a coherent state with $\alpha_\omega = -(\Gamma/\omega)(1-e^{i\omega t})$.
These results enable us to recover Eqs.~(\ref{commasympt0})-(\ref{commasympt}), as well as the contraction of the ensuing algebra, if one identifies $\gamma= 2 \pi \Gamma^2$, which is the Fermi golden rule.\footnote{We assumed that the bath is initially in its ground state $|0\rangle$. Different initial states are possible: for example, taking an initial thermal state of the bath would yield a different GKSL equation and a different decay rate $\gamma$, proportional to the number of thermal photons.}
This derivation offers a interesting perspective on the products~(\ref{prodpauli})-(\ref{anticommpauli}) and their limits~(\ref{prodpauliinf})-(\ref{anticommpauliinf}). As emphasized in Sec.~\ref{sec:intro2}, the limit captures, through the contraction, a symptom of irreversibility that is not detected for any finite $t$. Clearly, this can occur only with an infinite-dimensional dissipative bath.

The general features discussed for this particular simple model will be unaltered for other dissipative dynamical systems. We shall discuss other examples in the following.

\subsubsection{Phase damping of a harmonic oscillator.}
\label{sec:firsta}

Let
\begin{equation}
\label{oscphdampnew}
   L [\varrho] =  -\frac{\gamma}{2} \big(\{N^2,\varrho \} -2 N\varrho N \big),
\end{equation}
that describes a harmonic oscillator undergoing phase damping. Since $L^\sharp=L$ and $\Lambda^\sharp=\Lambda$, one finds
\begin{eqnarray}
\label{oscphdampt2}
\Lambda^\sharp_t [a] = e^{-\gamma t/2} a , \quad \Lambda^\sharp_t [a^\dagger] = e^{-\gamma t/2} a^\dagger ,  \quad \Lambda^\sharp_t [N] = N .
\end{eqnarray}
Note also that $\Lambda^\sharp_t [a^\dagger] \circ \Lambda^\sharp_t [a]= e^{-\gamma t} a^\dagger a$, so that
\begin{equation}
a^\dagger \circ_t a = (\Lambda^\sharp_t)^{-1} (\Lambda^\sharp_t [a] \circ \Lambda^\sharp_t [a])= e^{-\gamma t} a^\dagger a \to 0 .
\label{adaggera}
\end{equation}
Hence 
\begin{equation}
a^\dagger \circ_\infty a = 0 \quad \Rightarrow \quad a^\dagger = a = 0,
\label{avanish}
\end{equation}
in agreement with Eq.~(\ref{oscphdampt2}). Observe that (nonvanishing) $N$ cannot be identified with (vanishing) $a^\dagger \circ_\infty a$. This leaves us with the Abelian algebra  generated by $\{\mathbf{1} , N \}$, similarly to the example of the
phase damping of a qubit discussed in Sec.~\ref{sec:first}, and is consistent with physical interpretation: a generic density matrix
\begin{equation}
\label{rho0}
\varrho =  \sum c_{mn}  |m\>\<n| \; \stackrel{t \to \infty}{\longrightarrow} \; \varrho_\infty = \sum |c_{n}|^2  |n\>\<n|
\end{equation}
becomes diagonal in the $N$-representation, so that populations do not change and the ladder operators $a^\dagger$ and $a$ must vanish (weakly) over the final state.

As in the example of Sec.~\ref{sec:first}, we observe that the contraction of the original Heisenberg-Weyl oscillator algebra $\mathfrak{h}_4$ would yield the Lie algebra $\mathfrak{iso}(1,1)$ of the Poincar\'e group in $1+1$ dimensions,
\begin{equation}
[a,a^\dagger]_{\infty} =  0, \qquad [a,N]_{\infty}   =  a, \qquad [a^\dagger, N]_{\infty}  =-a^\dagger ,
\end{equation}
as can be checked by direct calculation. This is consistent with the Abelian algebra generated by $\{\mathbf{1} , N \}$ if Eq.~(\ref{avanish}) is taken into account.

\subsubsection{Comparison of the first two examples.}
\label{comparison}

A simple heuristic comparison between the first two examples shows that by the following substitution:
\begin{eqnarray}
a\rightarrow(\sigma_1-i\sigma_2)/2=\sigma_-,
\quad N\rightarrow(\sigma_0+\sigma_3)/2,
\end{eqnarray}
the commutation relations will be preserved and the equation of motion of the
example in Sec.~\ref{sec:firsta} will change into the equation of motion of the example in Sec.~\ref{sec:first}---up to the rescaling factor $1/2$.
The physical content of the two examples is, therefore, the same. However, mathematically this is only an analogy and should be taken with care, as the two algebras are different.
The physical analogy discussed here will be valid for all the examples that follows and it is a handwaving way to translate results obtained for qubits into analogous results for harmonic oscillators and \emph{vice versa}.

\subsection{Energy damping}

\subsubsection{Energy damping of a qubit.}
\label{sec:seconda}

Let
\begin{eqnarray}
L[\varrho]=-\frac{\gamma}{2}\big(\{\sigma_+ \sigma_-,\varrho\}-2\sigma_-\varrho\sigma_+\big).
\label{Ldec}
\end{eqnarray}
Hence, we have
\begin{eqnarray}
L^\sharp [A]=-\frac{\gamma}{2}\big(\{\sigma_+ \sigma_-,A\}-2\sigma_+A\sigma_-\big).
\end{eqnarray}
Observe that the evolution is not self-dual: $L^\sharp \neq L$.
We obtain
\begin{eqnarray}
\Lambda^{\sharp}_t[\sigma_{1,2}]=e^{-\gamma t/2}\sigma_{1,2},~~~\Lambda^{\sharp}_t[\sigma_{3}]=e^{-\gamma t}(\sigma_{3}+\sigma_0)-\sigma_0,~~~\Lambda^{\sharp}_t[\sigma_{0}]=\sigma_{0}
\end{eqnarray}
and
\begin{eqnarray}
\Lambda^{\sharp}_{\infty}[\sigma_{1,2}]=0,~~~\Lambda^{\sharp}_{\infty}[\sigma_{3}]=-\sigma_0,~~~\Lambda^{\sharp}_{\infty}[\sigma_{0}]=\sigma_{0}.
\end{eqnarray}
Observe also that $\sigma_{3} \circ_{\infty} \sigma_{3} =  \sigma_0$.
Thus the algebra is contracted to the one-dimensional Abelian algebra generated by the single element $\sigma_0$. 
This is in accord with the physical intepretation. The solution of Eq.~(\ref{Ldec}) reads
\begin{equation}
\label{qubit3solbis}
   \varrho = \frac{1}{2} (\sigma_0 + \bm x \cdot \bm\sigma) \quad \stackrel{t \to \infty}{\longrightarrow} \quad \varrho_\infty = P_-,
\end{equation}
so that the final state is the projection $P_- = (\sigma_0 - \sigma_3)/2$ over the ground state.

\subsubsection{Energy damping of a harmonic oscillator.}
\label{sec:second}

The following energy damping scenario can be attributed to the process of direct photodetection. For this dynamics, we have
\begin{eqnarray}
\label{oscdamp}
L [\varrho] &=  -\frac{\gamma}{2} \left( \{a^\dagger a,\varrho \} -2 a\varrho a^\dagger \right),
\end{eqnarray}
whence
\begin{eqnarray}
L^\sharp [A] &=  -\frac{\gamma}{2} \left( \{a^\dagger a,A \} -2 a^\dagger A a \right),
\end{eqnarray}
and
\begin{eqnarray}
\label{oscdampt2}
& & \Lambda^\sharp_t [a] = e^{-\gamma t/2} a , \quad \Lambda^\sharp_t [a^\dagger] = e^{-\gamma t/2} a^\dagger , \quad  \Lambda^\sharp_t [N] = e^{-\gamma t} N \quad (N=a^\dagger a).
\end{eqnarray}
Only unity survives the contraction.
Thus, similarly to the qubit example in Sec.~\ref{sec:seconda}, the oscillator algebra is contracted to the trivial Abelian algebra $\cA_\infty$ made up of a single element (unity) $\mathbf{1}$. This is consistent with physical interpretation, as the system decays to the ground state.

\subsection{Interaction with a thermal field}
\label{sec:more}

\subsubsection{Two-level atom in a thermal field.}
\label{sec:decay2}

In this example,
\begin{equation}
\label{decay2}
   L [\varrho] =  \frac{\gamma}{2} (n+1) (2 \sigma_-\varrho \sigma_+ - \{\sigma_+\sigma_-, \varrho\})
    + \frac{\gamma}{2} n (2 \sigma_+\varrho \sigma_- - \{\sigma_-\sigma_+, \varrho\} ),
\end{equation}
where $\sigma_\pm=(\sigma_1 \pm \sigma_2)/2$ and $n=(e^{\beta \hbar \Omega} -1)^{-1}$, with $\beta$ the inverse temperature and $\Omega$ the energy difference of the two states of the qubit. This dynamics generalizes the example in Sec.~\ref{sec:seconda} for $n\neq0$.

It can be easily checked that $L^{\sharp}[\sigma_3]=-2\gamma (n+\frac{1}{2})\sigma_3-\gamma \sigma_0$ and $L^{\sharp}[\sigma_0]=0$. Hence, $\sigma_3+\frac{1}{1+2n}\sigma_0$ is an eigen-operator of $L^{\sharp}$, $L^{\sharp}[\sigma_3+\frac{1}{1+2n}\sigma_0]=-2\gamma (n+\frac{1}{2})(\sigma_3+\frac{1}{1+2n}\sigma_0)$, and the time evolutions of $\sigma_3$ and $\sigma_0$ are easily found to be
\begin{eqnarray}
\Lambda^{\sharp}_t[\sigma_3]=e^{-\gamma(1+2n)t}\sigma_3+\frac{1}{1+2n}(e^{-\gamma(1+2n)t}-1)\sigma_0, \quad\Lambda^{\sharp}_t[\sigma_0]=\sigma_0,
\end{eqnarray}
so that
\begin{eqnarray}
\Lambda^{\sharp}_\infty[\sigma_3]=-\frac{1}{1+2n}\sigma_0, \quad\Lambda^{\sharp}_\infty[\sigma_0]=\sigma_0.
\label{asyms3}
\end{eqnarray}
The observables $\sigma_1$ and $\sigma_2$ are also eigen-operators of $L^\sharp$, since $L^\sharp[\sigma_{1,2}]=-\gamma(n+\frac{1}2)\sigma_{1,2}$, and their time evolutions are
\begin{equation}
\Lambda^{\sharp}_t[\sigma_{1,2}]=e^{-\gamma(n+1/2)t}\sigma_{1,2} \to \Lambda^{\sharp}_\infty[\sigma_{1,2}] =0 .
\label{asyms12}
\end{equation}
The contracted algebra $\cA_\infty$ is Abelian and is generated by the single element $\sigma_0$. This is in accord with the asymptotic solution of Eq.~(\ref{decay2}),
\begin{equation}
\label{qubitthsol}
   \varrho = \frac{1}{2} (\sigma_0 + \bm x \cdot \bm\sigma) \stackrel{t \to \infty}{\longrightarrow}  \Lambda_\infty[\varrho]=\varrho_\infty = \frac{P_+ + e^{-\beta \hbar \Omega} P_-}{1+ e^{-\beta \hbar \Omega}} ,
\end{equation}
where the notation is as in Eq.~(\ref{qubit3sol}), and $P_\pm = (\sigma_0 \pm \sigma_3)/2$ are the two projections.

Direct computation of the Lie algebra enables one to make a further remark.
One obtains
\begin{eqnarray}
&[\sigma_1,\sigma_2]_{\infty}=2i (\sigma_3+\frac{1}{1+2n}\sigma_0) , \label{rhs0}
\\
& [\sigma_2,\sigma_3]_{\infty}=0,\\
&[\sigma_3,\sigma_1]_{\infty}=0.
\end{eqnarray}
Curiously, we encounter a central extension of the Heisenberg-Weyl algebra $\mathfrak{h}_3$ ($a,a^\dagger,\mathbf{1}$, without $N$). However, interestingly, Eq.~(\ref{asyms12}) forces the right-hand side of Eq.~(\ref{rhs0}) to vanish. For consistency, the left-hand side must vanish too. This yields
\begin{eqnarray}
\sigma_3+\frac{1}{1+2n}\sigma_0 \; \propto \; P_+ - e^{-\beta \hbar \Omega} P_- =0,
\end{eqnarray}
which is Boltzmann's statistics.

%%%%%%%%%%%%%%%%%%%%%%%%%%%%%%%%%%%%%%%%%%%%%%%%%%%%%%%%%%%%%%%%%%%%%%%%%%
\subsubsection{Thermal damping of a harmonic oscillator.}
\label{sec:damping}

Let
\begin{equation}
\label{damp}
   L [\varrho] =  \frac{\gamma}{2} (m+1) (2 a\varrho a^\dagger - a^\dagger a \varrho - \varrho a^\dagger a)
    + \frac{\gamma}{2} m (2 a^\dagger\varrho a - a a^\dagger \varrho - \varrho a a^\dagger) .
\end{equation}
Here again, $m=(e^{\beta \hbar \Omega} -1)^{-1}$, with $\beta$ inverse temperature and $\Omega$ the oscillator frequency.
The solution is
\begin{eqnarray}
&\Lambda^{\sharp}_t [a]=e^{-\frac{\gamma}{2}t}a,~~~\Lambda^{\sharp}_t [a^{\dag}]=e^{-\frac{\gamma}{2}t}a^{\dag}.
\end{eqnarray}
Since $L^{\sharp}[N]=-\gamma (N-m \mathbf{1})$ and $L^{\sharp}[\mathbf{1}]=0$, one easily verifies that $N- m \mathbf{1}$ is an eigen-operator of $L^{\sharp}$
with eigenvalue $-\gamma$. Thus, the time evolution of this operator can be easily obtained as 
\begin{eqnarray}
&\Lambda^{\sharp}_t [N- m \mathbf{1}]=e^{-\gamma t}(N-m \mathbf{1}),
\end{eqnarray}
whence the time evolution of $N$ becomes
\begin{eqnarray}
&\Lambda^{\sharp}_t [N]=e^{-\gamma t}N+m(1-e^{-\gamma t})\mathbf{1}.
\end{eqnarray}
From this we conclude that
\begin{eqnarray}
\Lambda^{\sharp}_{\infty} [a]=0,~~~\Lambda^{\sharp}_{\infty} [a^{\dag}]=0,~~~\Lambda^{\sharp}_{\infty} [N]=m \mathbf{1},
\end{eqnarray}
which implies that the contracted algebra $\cA_\infty$ is Abelian and made up of a single element (unity): $\mathbf{1}$.
As in the example of Section~\ref{sec:decay2}, the final state is thermal and the result is consistent.

\subsection{Two-level atom in a squeezed vacuum}
\label{sec:sueeze}

Let
\begin{eqnarray}
   L [\varrho] & = & \frac{\gamma}{2} (n+1) (2 \sigma_-\varrho \sigma_+ - \{\sigma_+\sigma_-, \varrho\})
    + \frac{\gamma}{2} n (2 \sigma_+\varrho \sigma_- - \{\sigma_-\sigma_+, \varrho\} ) \nonumber  \\
    & &  - \ \gamma (m \sigma_+\varrho \sigma_+ + m^* \sigma_-\varrho \sigma_-), \label{squeeze}
\end{eqnarray}
where $|m|^2 \leqslant n(n+1)$.
Since $\sigma_3$ and $\sigma_0$ are the eigen-operators of the third term of $L^{\sharp}$ with eigenvalue $0$, the time evolution of these operators is not different from that found in the previous example (where $m$ was zero). That is,
\begin{eqnarray}
\Lambda^{\sharp}_t[\sigma_3]=e^{-\gamma(1+2n)t}\sigma_3+\frac{1}{1+2n}(e^{-\gamma(1+2n)t}-1)\sigma_0, \quad \Lambda^{\sharp}_t[\sigma_0]=\sigma_0.
\end{eqnarray}
It can be easily calculated that
\begin{equation}
L^{\sharp}[\sigma_1]=-\gamma\Big((n+1/2)+(m+m^*)/2\Big)\sigma_1-i\gamma (m-m^{*})\sigma_2/2,
\end{equation}
whence
\begin{equation}
L^{\sharp}[\sigma_2]=-\gamma\Big((n+1/2)-(m+m^*)/2\Big)\sigma_2-i\gamma (m-m^{*})\sigma_1/2.
\end{equation}
After some algebra, one can find that $\sigma_2-i\frac{m-m^*}{m+m^{*}\pm 2|m|}\sigma_1$ is an eigen-operator of $L^{\sharp}$ with eigenvalue $-\gamma(n+1/2)\pm \gamma |m|$. Hence, we obtain the time evolutions of these two eigen-operators as
\begin{eqnarray}
\Lambda^{\sharp}_t[\sigma_2-i\frac{m-m^*}{m+m^{*}+2|m|}\sigma_1]&=e^{-\gamma(n+|m|+1/2)t}(\sigma_2-i\frac{m-m^*}{m+m^{*}+ 2|m|}\sigma_1), \nonumber \\
\\
\Lambda^{\sharp}_t[\sigma_2-i\frac{m-m^*}{m+m^{*}-2|m|}\sigma_1]&=e^{-\gamma(n-|m|+1/2)t}(\sigma_2-i\frac{m-m^*}{m+m^{*}- 2|m|}\sigma_1). \nonumber \\
\end{eqnarray}
Thus the time evolutions of $\sigma_{1,2}$ are obtained as
\begin{eqnarray}
\Lambda^{\sharp}_t[\sigma_{1}] &=& e^{-\gamma(n+1/2)t}\big[\sigma_1\{\frac{m+m^{*}}{2|m|}\sinh(\gamma |m|t)+\cosh(\gamma |m|t)\} \nonumber \\
& & +\sigma_2\{i\frac{m-m^{*}}{2|m|}\sinh(\gamma |m|t)\}\big]\\
\Lambda^{\sharp}_t[\sigma_{2}] &=& e^{-\gamma(n+1/2)t}\big[\sigma_2\{-\frac{m+m^{*}}{2|m|}\sinh(\gamma |m|t)+\cosh(\gamma |m|t)\} \nonumber \\
& &
+\sigma_1\{i\frac{m-m^{*}}{2|m|}\sinh(\gamma |m|t)\}\big]
\end{eqnarray}
Having $|m|^2 \leqslant n(n+1)$ we find $n\pm|m|+1/2>0$. As a result, when $t$ goes to infinity we obtain
\begin{eqnarray}
\Lambda^{\sharp}_{\infty}[\sigma_3]&=-\frac{1}{1+2n}\sigma_0, ~~~\Lambda^{\sharp}_{\infty}[\sigma_0]=\sigma_0,\\
\Lambda^{\sharp}_{\infty}[\sigma_{1}]&=0,~~~ \Lambda^{\sharp}_{\infty}[\sigma_{2}]=0 .
\end{eqnarray}
This is the same algebra as in the previous examples. The same considerations apply too.

\section{$N$-level quantum system.}
\label{sec:gener}

The examples in the preceding section enabled us to familiarize with the contracted product and its physical implications. We can now generalize some salient features of the technique by looking at a simple example involving an $N$-level quantum system.

Consider  the following construction: let  $\mathcal{P}$ be  a completely-positive trace preserving projection, i.e., $\mathcal{P}^2 = \mathcal{P}$, and let  $\mathcal{P}^\perp = {\rm id} - \mathcal{P}$ (${\rm id}$ denotes the identity map, i.e., ${\rm id} [A]= A$ for all $A \in \cA$). Note that $\mathcal{P}^\perp$ contrary to $\mathcal{P}$ is not completely positive. Now, define the generator
\begin{equation}
\label{generator}
    L  = - \gamma \mathcal{P}^\perp \ ,
\end{equation}
with $\gamma >0$.  One easily finds the evolution
\begin{equation} 
    \Lambda_t = \mathcal{P} + e^{-\gamma t} \mathcal{P}^\perp\ .
\end{equation}
Hence the asymptotic dynamical map reads as $\Lambda_\infty = \mathcal{P}$. The inverse map reads as
\begin{equation} 
    \Lambda_t^{-1} = \mathcal{P} + e^{\gamma t} \mathcal{P}^\perp\ .
\end{equation}
and hence one finds the corresponding limiting product 
\begin{eqnarray}
 A \circ_\infty B & = & \mathcal{P}^\sharp[\mathcal{P}^\sharp [A] \circ \mathcal{P}^\sharp [B] ] 
 \nonumber \\
 & & + \mathcal{P}^{\perp \sharp} \left[ \mathcal{P}^{\perp\sharp} [A] \circ \mathcal{P}^\sharp [B] + \mathcal{P}^\sharp [A] \circ \mathcal{P}^{\perp\sharp} [B] \right] , 
 \label{P-product}
\end{eqnarray}
for any $A,B \in \mathcal{A}$. This formula shows that $(\mathcal{A}_\infty,\circ_\infty)$ defines an associative algebra. Indeed, if $A,B \in \mathcal{P}^\sharp[\cA]$, then $ A \circ_\infty B \in \mathcal{P}^\sharp[\cA]$. Moreover, if $\oper$ denotes the unit element in $\cA$, then $\mathcal{P}^\sharp  [\oper] = \oper$, and hence
$\oper \circ_\infty A = A$, for all $A \in \cA$.

The asymptotic algebra is defined by the projection $\mathcal{P}$, that is, $\mathcal{A}_\infty = \mathcal{P}^\sharp[\mathcal{A}]$. Hence, we can freely model the asymptotic algebra by choosing an appropriate projection $\cal P$.

Let us consider a completely-positive projector in $\mathcal{A} = \mathcal{B}(\mathcal{H})$ represented as follows\begin{color}{red}:\end{color}
\begin{equation} 
  \mathcal{P}^\sharp [A] = \sum_k P_k A P_k ,
\end{equation}
where $P_k$ is a family of mutually orthogonal projectors in $\mathcal{H}$ such that $\sum_k P_k = \oper$. Note that $\mathcal{P}^\sharp = \mathcal{P}$.
One can easily check
\begin{equation} 
  \mathcal{P}^\sharp [ \mathcal{P}^\sharp[A] \circ B] =  \mathcal{P}^\sharp[ A \circ \mathcal{P}^\sharp[B]] = \mathcal{P}^\sharp[A] \circ  \mathcal{P}^\sharp[B] ,
\end{equation}
which finally implies
\begin{equation} 
   A \circ_\infty B = \mathcal{P}^\sharp [A] \circ \mathcal{P}^\sharp [B]  +  \mathcal{P}^{\perp\sharp} [A] \circ \mathcal{P}^\sharp [B] + \mathcal{P}^\sharp [A] \circ \mathcal{P}^{\perp\sharp} [B] ,
\end{equation}
and hence
\begin{equation}\label{X}
  A \circ_\infty B = A \circ B - \mathcal{P}^{\perp\sharp} [A] \circ \mathcal{P}^{\perp\sharp} [B]\ .
\end{equation}
In fact, we have the following proposition:
\begin{Proposition} 
$A \in \mathcal{A}_{\infty}$ iff
\begin{equation} 
  A \circ_\infty B = A \circ B ,
\end{equation}
for all $B \in \mathcal{A}$.
\end{Proposition}
Indeed, due to Eq.~(\ref{X})$, A \circ_\infty B = A \circ B $ for all $B \in \mathcal{A}$ if and only if $\mathcal{P}^{\perp\sharp} [A] =0$ which means that $A \in \mathcal{A}_\infty$.

Let us look at two simple examples. Taking the Hilbert space $\mathcal{H} = \mathbb{C}^2$ and
\begin{equation} 
    \mathcal{P}[\varrho] = P_+ \varrho P_+ + P_- \varrho P_-\ ,
\end{equation}
with $P_\pm = (\sigma_0 \pm \sigma_3)/2$, one reconstructs Eq.\ (\ref{qubit3}). 
In fact, one obtains the following non-commutative deformed matrix multiplication \cite{cgm2}
\begin{equation}
\label{NEW}
\left( \begin{array}{cc} a_{11} & a_{12} \\ a_{21} & a_{22}\end{array} \right) \cdot_\infty
    \left( \begin{array}{cc} b_{11} & b_{12} \\ b_{21} & b_{22}\end{array} \right) =
    \left( \begin{array}{cc} a_{11}b_{11} & a_{11}b_{12}+a_{12}b_{22} \\ a_{21}b_{11}+ a_{22}b_{21}& a_{22}b_{22}\end{array} \right),
\end{equation}
namely, 
\begin{equation}\label{HAD}
    (A \cdot_\infty B)_{ij} = \left\{ \begin{array}{ll} (A \cdot B)_{ij} \ & \ i\neq j \\ (A \cdot_h B)_{ij} \ & \ i=j \end{array} \right.  ,
\end{equation}
where $A \cdot_h B$ denotes the Hadamard product.

Taking $\mathcal{H} = \mathbb{C}^3$ and
\begin{equation} 
    \mathcal{P}[\varrho] = P \varrho P + P^\perp \varrho P^\perp\ ,
\end{equation}
with $P$ one-dimensional and $P^\perp$ two-dimensional, one has the following example.
Consider a 3-level system $\{|a\>,|b\>,|c\>\}$ and let
\begin{equation}
\label{decfree2}
   L [\varrho] =   \gamma (2P^\perp \varrho P^\perp - P^\perp \varrho - \varrho P^\perp),
\end{equation}
where $P^\perp=|b\>\<b|+ |c\>\<c|$.
It is straightforward to check that
\begin{eqnarray}
\label{rabc}
& & |r\>\<r| \;  \stackrel{t \to \infty}{\longrightarrow}  \; |r\>\<r| \qquad (r=a,b,c) \\
& & |a\>\<r| \pm  |r\>\<a| \;  \stackrel{t \to \infty}{\longrightarrow}  \;  0
\label{aroff} \qquad (r=b,c) \\
& & |b\>\<c| \pm  |c\>\<b| \;  \stackrel{t \to \infty}{\longrightarrow}  \;  |b\>\<c| \pm  |c\>\<b|
\label{bcoff}
\end{eqnarray}
Equation~(\ref{rabc}) guarantees population preservation in every level,
Eq.~(\ref{aroff}) hinders transitions between level $a$ and the other two levels, whereas 
Eq.~(\ref{bcoff}) shows that the dynamics within the subspace $\mathrm{span}\{|b\>,|c\>\}$ is preserved.
Note that the original Lie algebra is $\cA=\mathfrak{su}(3)$, while the final algebra $\cA_\infty =  \Lambda^\sharp_\infty [\cA]$ contains $\mathfrak{su}(2)$ as a subalgebra (on $\mathrm{span}\{|b\>,|c\>\}$). Summarizing, for a generic density matrix
\begin{equation}   \label{ppp}
\varrho  = \left(
\begin{array}{ccc}
p_a & \varrho_{ab} & \varrho_{ac} \\
\varrho_{ab}^* & p_b & \varrho_{bc} \\
\varrho_{ac}^* & \varrho_{bc}^* & p_c
\end{array}
\right)
\; \longrightarrow \;
\varrho_\infty =
\left(\begin{array}{ccc}
p_a & 0 & 0 \\ 0 & p_b & \varrho_{bc} \\0 & \varrho_{bc}^* & p_c \end{array}\right),
\end{equation}
and
\begin{eqnarray} 
 && \left(
\begin{array}{ccc}
A_{aa} & A_{ab} & A_{ac} \\
A_{ba} & A_{bb} & A_{bc} \\
A_{ca} & A_{cb} & A_{cc}
\end{array}
\right) \circ_\infty  \left(
\begin{array}{ccc}
B_{aa} & B_{ab} & B_{ac} \\
B_{ba} & B_{bb} & B_{bc} \\
B_{ca} & B_{cb} & B_{cc}
\end{array}
\right) = \nonumber \\ && \quad =  \left(
\begin{array}{ccc}
A_{aa} & A_{ab} & A_{ac} \\
A_{ba} & A_{bb} & A_{bc} \\
A_{ca} & A_{cb} & A_{cc}
\end{array}
\right) \circ  \left(
\begin{array}{ccc}
B_{aa} & B_{ab} & B_{ac} \\
B_{ba} & B_{bb} & B_{bc} \\
B_{ca} & B_{cb} & B_{cc}
\end{array}
\right) \nonumber \\ && \quad \qquad - \left(
\begin{array}{ccc}
0 & A_{ab} & A_{ac} \\
A_{ba} & 0 & 0 \\
A_{ca} & 0 & 0
\end{array}
\right) \circ  \left(
\begin{array}{ccc}
0 & B_{ab} & B_{ac} \\
B_{ba} & 0 & 0 \\
B_{ca} & 0 & 0
\end{array}
\right) \ ,
\end{eqnarray}
that generalizes the product (\ref{NEW}).
The dynamics between levels $|b\>$ and $|c\>$ is unitary and the asymptotic algebra reflects the underlying quantum coherence.

\section{Conclusions and outlook.}
\label{sec:concl}

One of the main differences between classical and quantum systems lies in the commutation properties of its observables. Noncommutativity is a distinctive quantum feature: observables that can be simultaneously measured are ``classical" with respect to each other, and when all observables commute the system can be viewed as fully classical. These notions can be formulated in terms of the algebra of the operators of the system, and therefore can be traced back to the structure of the associative product among them.

Dissipation and decoherence always tend to deteriorate quantum features of a system and make it increasingly classical. In this article, we have defined a product that detects the dissipative features of the evolution and the
increasing difficulty in measuring those observables that are more affected by decoherence and dissipation. In the long-time limit, this procedure yields a contracted algebra of operators. The contracted algebra always contains commutative subalgebra(s) and is, therefore, more ``classical" than the original one.

The whole procedure is based on Dirac's prescription~(\ref{SvsH}). We have worked out a number of examples, and have seen that a key role is played by the asymptotic state of the evolution.

In this framework, ample room is left for noncommutative (quantum) observables, that do not belong to the center of the contracted algebra. These are associated with the kernel of $L^\sharp$, which coincides with the commutant of the interaction algebra, that is the algebra generated by the Kraus
operators and their adjoints \cite{kribs,BN,ZCV}. These observables are not affected by dissipation and preserve their quantum features. We have analyzed in detail a paradigmatic case study for the dissipative evolution of an $N$-level quantum system. In general, the generator in Eq.~(\ref{generator}) are determined by the physics of the problem. A part of the initial algebra will survive the contraction and will define those observables that can be measured on the quantum dissipative system.

The contractions bear the consequences of the irreversibility of the dynamics. We will discuss in a future article the extension of the present framework to more general non-unitary evolutions, such as quantum channels.

\ack
This work was partially supported by PRIN 2010LLKJBX on ``Collective quantum phenomena: from strongly correlated systems to quantum simulators,'' and by the Italian National Group of Mathematical Physics (GNFM-INdAM).

\end{document}